\documentclass[preprint,1p,review]{elsarticle}

\usepackage{amssymb,amsmath,array,adjustbox,xcolor,bm}
\usepackage{graphicx}
\usepackage{epstopdf}
\usepackage[linesnumbered,ruled] {algorithm2e}
\usepackage{nomencl}
\usepackage{algpseudocode}

\usepackage{etoc,longtable,caption}
\usepackage[hidelinks]{hyperref}
\newcolumntype{C}[1]{>{\arraybackslash}p{#1}}

\journal{Engineering Applications of Artificial Intelligence}

\begin{document}
	
	\begin{frontmatter}
		
		\title{\qquad Robust Auto-landing Control of an agile Regional Jet Using Fuzzy Q-learning}

		\author[1]{Mohsen Zahmatkesh}
		
		\author[1]{Seyyed Ali Emami}

		\author[1]{Afshin Banazadeh \corref{mycorrespondingauthor}}
		\cortext[mycorrespondingauthor]{Corresponding author}
		\ead{banazadeh@sharif.edu}
		
		\author[2]{Paolo Castaldi}
		
		\address[1]{Department of Aerospace Engineering, Sharif University of Technology, Tehran, Iran}
		\address[2]{Department of Electrical, Electronic and Information Engineering "Guglielmo Marconi", University of Bologna, Via Dell’Universit`a 50, Cesena, Italy}
		
	\begin{abstract}                
A robust auto-landing problem of a Truss-braced Wing (TBW) regional jet aircraft with poor stability characteristics is presented in this study employing a Fuzzy Reinforcement Learning scheme. Reinforcement Learning (RL) has seen a recent surge in practical uses in control systems. In contrast to many studies implementing Deep Learning in RL algorithms to generate continuous actions, the methodology of this study is straightforward and avoids complex neural network architectures by applying Fuzzy rules. An innovative, agile civil aircraft is selected not only to meet future aviation community expectations but also to demonstrate the robustness of the suggested method. In order to create a multi-objective RL environment, a Six-degree-of-freedom (6-DoF) simulation is first developed. By transforming the auto-landing problem of the aircraft into a Markov Decision Process (MDP) formulation, the problem is solved by designing a low-level Fuzzy Q-learning (FQL) controller. More specifically, the well-known Q-learning method, which is a discrete RL algorithm, is supplemented by Fuzzy rules to provide continuous actions with no need to complex learning structures. The performance of the proposed system is then evaluated by extensive flight simulations in different flight conditions considering severe wind gusts, measurement noises, actuator faults, and model uncertainties. Besides, the controller effectiveness would be compared with existing competing techniques such as Dynamic Inversion (DI) and Q-learning. The simulation results indicate the superior performance of the proposed control system as a reliable and robust control method to be employed in real applications.
	\end{abstract}
	
	
\end{frontmatter}

\section{Introduction}
Recently, the aviation industry has faced a number of challenges, including increased emissions and congested airspace. As a result, the future world requirements will be characterized as safety and efficiency. In the efficiency segment, reducing emissions by using less fuel and making economical flights alongside faster flights is the main goal. On another hand, high-speed flight management in moreover congested airspace falls within the subject of safety.
In conclusion, the discussed parameters strongly encourage the development of novel aircraft configurations to gain advantageous elements. The Scope Clause, on the other hand, is an agreement that places a cap on the number of aircraft seats in order to prevent outsourcing and guarantee the jobs of union pilots. Therefore, it is inevitable that the Modern Regional Jet (MRJ) fleet would grow. That demonstrates the importance of reliable flight control systems. In high-performance aerodynamic configurations, such as TBW aircraft, there are some re-raised interests \cite{li2022multipoint,sohst2022optimization,chau2022aerodynamic}. Apart from that, the aviation industry has expressed interest in TBWs because of their fuel burn efficiency \cite{zavaree2021modern}. Despite considerable dynamic modeling research \cite{nguyen2022dynamic}, it appears that no reliable auto-landing controller has been developed for these configurations.

Studies show that the landing procedure is the riskiest part of flying and calls for expert pilotage abilities. The International Civil Aviation Organization (ICAO) 6th annex document introduces three alternate aerodromes—Take-off, En-route, and Destination—where an MRJ shall land in the event of particular failure scenarios. In this case, \cite{Suharev2019AnalysisOT} examined the most frequent causes of flying incidents and accidents during the approach phase. The most hazardous elements included are pilot erroneous decisions (74\%), skipping or completing activities incorrectly (72\%), and ineffective crew communication, mutual cooperation, and mutual control (63\%). The cited elements strengthen the function of trendy autonomous controllers with great dependability to bolster safety.
There are several studies with advanced control centralization in the landing procedure. For instance, \cite{EROGLU2020105855} developed a double-loop nonlinear dynamic inversion controller based on a deep failure estimator made up of layers of Long-short Term Memory (LSTM) and Convolution Neural Networks (CNN). These neural networks were trained for severe stuck failures, with time-series data on landing trajectory patterns in actuator faulty and healthy landing condition simulations. In order to control the trajectory of a tailless and blended wing UAV confronting air turbulences and sensor measurement errors, a unique auto-landing framework was presented in \cite{LUNGU2022107261}. In this research, a Backstepping-based controller is used to control the attitude angles. A Dynamic Inversion-based controller creates the throttle signal to keep a constant velocity, and an adaptive disturbance observer estimates the atmospheric turbulences to track the proposed landing trajectory. In \cite{ISMAIL201455} a generalized Anti-windup based on traditional PID controllers as well as a phase compensation system used to train a neural-assisted Sliding Mode controller. In order to tackle stuck actuator failures during an auto-landing scenario. It is demonstrated that the capacity of the neural-aided Sliding Mode controller to tolerate faults is greatly improved by adding Anti-windup and phase compensation. An aircraft heading angle is guided by a Deep Q-learning (DQL) in \cite{8695548}, allowing it to land in the desired 2-dimensional field. In this study, the dynamic modeling of aircraft is not included. In \cite{8866189} a Deep Deterministic Policy Gradient (DDPG) was used to control an Unmanned Aerial Vehicle (UAV's) desired path for the landing flare phase in the longitudinal channel in existing wind disturbances. The structure of the proposed method includes two Deep ANNs as an Actor-critic architecture. Similar to this, the DDPG approach is employed in the outer loop of a landing procedure in \cite{9213987} whereas the inner loop is controlled by a Proportional-integral-derivative (PID) controller. It is important to note that several classic techniques for aircraft attitude control have been created to improve the quality of landings. The main shortcoming of these traditional theories is their lack of adaptation in extended working points. To overcome this weakness, some existing approaches have increased their robustness by utilizing ANNs.

In this regard, publications like \cite{liao2005fault} and \cite{PASHILKAR200649} are considered. In the first paper, a $H_2$ controller is addressed to actuator faults and wind disturbances. The focus of the second paper is on designing an online neural-aided controller to increase the robustness of existing controllers for fault tolerance.
Although not specified in the landing phase, a different group of publications also deals with controller design. For instance, \cite{zogopoulos2021fault} presented a layered Model Predictive Control system based on simple sparse rapidly exploring random trees for path planning, path control, velocity control, and angular velocity control. This method improved the estimation of a real-time, nonlinear, and onboard convex Flight-envelop calculation. For hydraulic actuator failures, \cite{IJAZ20191302} created an Integral Sliding Mode controller featuring Control Allocation. This controller eliminates the need to redesign another controller owing to distributing control signals among redundant actuators.
An ANN-based adaptive controller using Feedback Linearization was developed in \cite{doi:10.1177/0954410018758497} to address the dramatic roll and unsteady longitudinal behavior in the condition of partial wing damage. In another study, to achieve hydraulic fault tolerance in the longitudinal axis, \cite{norgaard2017performance} evaluated the performance of three controllers, including Adaptive Back-stepping, Robust Sliding Mode, and PID. In this instance, the initial controller overcame the other methods. A Q-learning horizontal trajectory tracking controller with an ANN foundation was developed in \cite{YANG2020106100} using the MDP model of an airship with fine stability characteristics. In this study, the method of action selection was optimized using a Cerebellar Model Articulation Controller (CMAC) neural network. A Soft Actor-critic (SAC) technique was used in \cite{xi2022energy} to solve a path planning problem for a Long-endurance, Solar-powered UAV that took energy consumption into account.
Another study cited as \cite{bohn2021data} focused on a Skywalker X8 inner loop control employing SAC and comparing it with a PID controller.

Proximal Policy Optimization (PPO), was used in \cite{hu2022fixed} for orientation control of a typical extremely dynamic coupled Fixed-wing aircraft in the stall circumstance. After 100,000 episodes, there was a successful convergence of the PPO. 
The efforts that have been mentioned thus far use ANNs to improve convergence and robustness. To the best of our knowledge, there are discrete RL-based attitude control studies without using ANNs. A Q-learning method was used in \cite{10.1007/978-3-030-98404-5_59} to control longitudinal and lateral angles in a general aviation airplane (Cessna 172). This research controls the desired angles of zero and the airplane profits good stability characteristics.

There are some Fuzzy adaptations on \cite{watkins1992q} work like \cite{622790} where the Q-functions and action selection strategy are inferred from Fuzzy rules. Also, in order to reduce the number of states needed to shape an MDP model for mobile robots that avoid obstacles, \cite{8834601} suggests a Fuzzy technique.
Because the mobile robot may encounter an infinite number of different conditions. The Fuzzy method is used to generalize the condition and minimize processor requirements alongside reducing states. Also, \cite{1298895} proposed a dynamic Fuzzy Q-learning for online and continuous tasks in mobile robots. In \cite{9789160}, the Fuzzy Q-learning (FQL) method and Strictly Negative Imaginary (SNI) property are used to provide a novel robust adaptive control for quadrotor attitude and altitude stabilization. The objective is to develop a control strategy that dynamically adapts the SNI controller using FQL. Another study, \cite{act11120374} used Q-learning as an attitude controller for a unique, highly maneuverable regional jet aircraft in MDP and POMDP scenarios. The simulation results in a variable pitch angle tracking were satisfactory.

\begin{figure}[h!]
	\begin{center}
		\includegraphics[width=10.5cm]{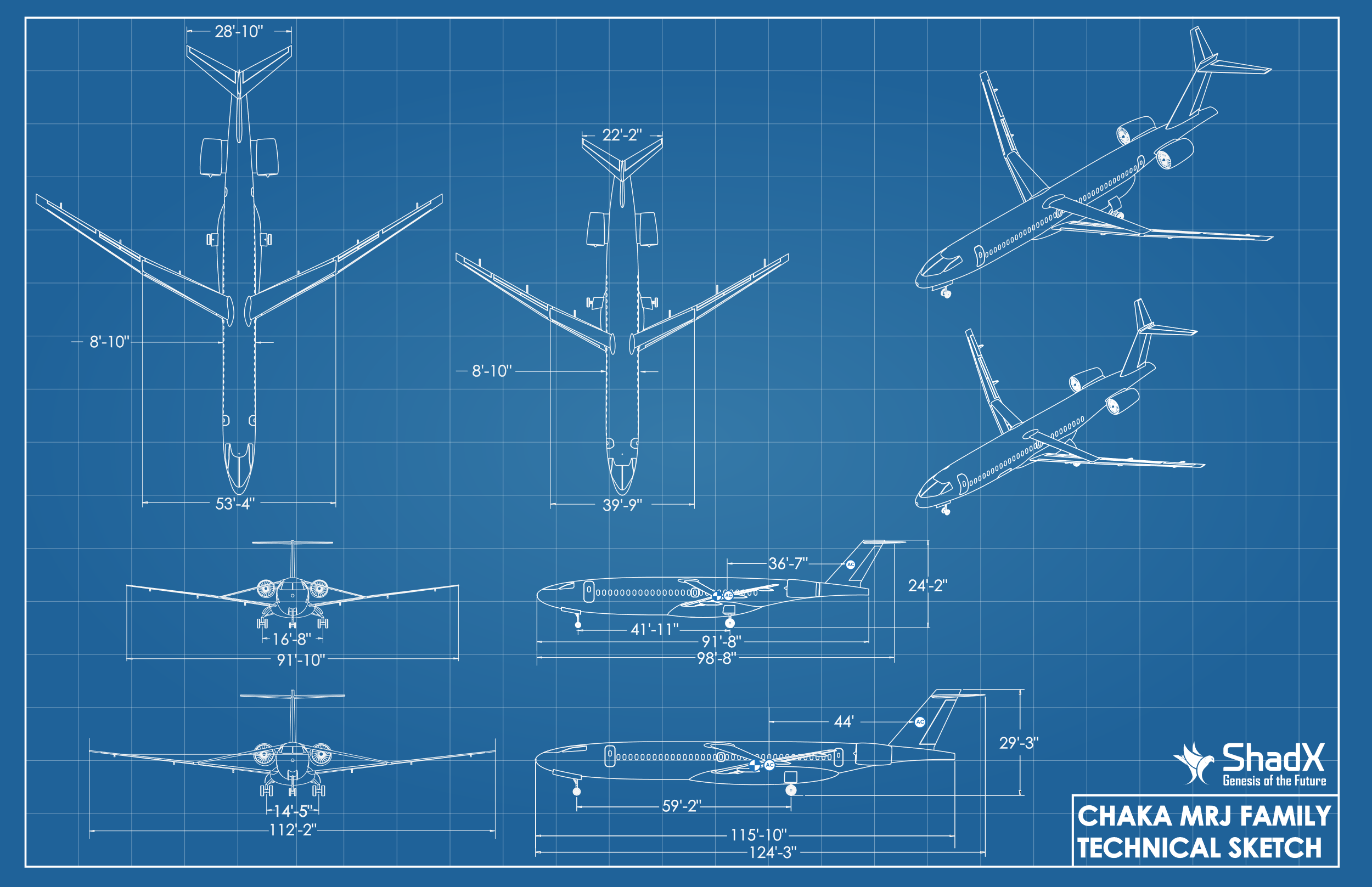}    
		\caption{Chaka 50 and 76 Modern Regional Jet (MRJ) Family \cite{zavaree2021modern}}
		\label{fig:1}
	\end{center}
\end{figure}

Motivated by the preceding discussions, the following are the main contributions of the current study: 
\begin{enumerate}
  \item A novel continuous action generator is developed as a general connector between every (discrete/continuous) optimal policy and the RL environment.
  \item In response to worldwide aviation community expectations, a TBW aircraft (figure \ref{fig:1}) with specific stability characteristics is selected for the auto-landing problem, where the high maneuverability of the aircraft brings significant challenges into the design process.
  \item In contrast to many studies, the complexity of ANN architectures and the low adaptation of classic methods are well resolved using Fuzzy Q-learning.
  \item The robustness and reliability of the proposed FQL are examined under different flight conditions consisting of sensor measurement noises, atmospheric disturbances, actuator faults, and model uncertainties.
  
\end{enumerate}

\section{Six-DoF Aircraft Dynamic Modeling}
In order to develop a nonlinear RL environment, the 6-DoF nonlinear equations of motions cited in \cite{napolitano2012aircraft, zipfel2014modeling} are utilized in this section.
Many environments based on Gym and Flight Gear are open-source, such as GymFG (\cite{wood2020gymfg}).
But this plant must model and simulate from scratch due to the unique characteristics of the innovative configuration.
In this approach, it is presumed that the earth is flat.
Consequently, the body frame translational and rotational equations are as follows:
\begin{equation}\label{Eq:1}
	mD^B \bm{v}^B+ m \bm{\Omega} ^{B} \bm{v}^B= \bm{f}_{a}^B + \bm{f}_{p}^B +m\bm{g}^B,
\end{equation}
\begin{equation}\label{Eq:2}
	D^B(\bm{I}^B \bm{\omega}^{B})+\bm{\Omega}^{B}\bm{I}^B\bm{\omega}^{B}=\bm{m}_a^B + \bm{m}_p^B.
\end{equation}
Where $u, v, w$ are the velocity components, and $D^B$ is defined as the Rotational Time Derivative in body frame. So the $D^B \bm{v}^B$ equals $[\frac{dv}{dt}]^B = [\dot{u}\ \ \dot{v}\ \ \dot{w}]^T$, and $[\omega]^B = [p\ \ q\ \ r]^T$ are the roll, pitch, and yaw angular rates vector in body frame. Also $\bm{\Omega}^B$ is the skew-symmetric form of angular rates vector.
\begin{equation}\label{Eq:3}
	[\Omega]^B = \begin{bmatrix}
				0 & -r & q \\
				r & 0 & -p \\
				-q & p & 0
			\end{bmatrix}.
\end{equation}
Furthermore, $m$ is the mass of aircraft, and $\bm{I}^B$ is the moment of inertia matrix in body frame;
\begin{equation}\label{Eq:4}
	[I]^B = \begin{bmatrix}
				I_x & 0 & I_{xz} \\
				0 & I_y & 0 \\
				I_{xz} & 0 & I_z
			\end{bmatrix}.
\end{equation}
In equations (\ref{Eq:1}) and (\ref{Eq:2}), some variables on the right-hand side are considered to be zero including $\bm{m}_p^B$ vector which is engine power moments. Also, $\bm{f}_{a}^B$, $\bm{f}_{p}^B$, and $\bm{m}_{a}^B$ are considered as aerodynamic force, engine power force, and aerodynamic moment vectors. Where except engine power, the other non-zero forces and moments are computed in aerodynamic frame as follows;
\begin{equation}\label{Eq:5}
\begin{aligned}
	\begin{bmatrix}
		L \\
		D \\
		m_a
	\end{bmatrix}^S =\bar{q} S \bar{c}
	\begin{bmatrix}
		c_{L_0} & c_{L_\alpha} & c_{L_{\dot{\alpha}}} & c_{L_u} & c_{L_q} & c_{L_{\delta_E}}\\
		c_{D_0} & c_{D_\alpha} & c_{D_{\dot{\alpha}}} & c_{D_u} & c_{D_q} & c_{D_{\delta_E}}\\
		c_{m_0} & c_{m_\alpha} & c_{m_{\dot{\alpha}}} & c_{m_u} & c_{m_q} & c_{m_{\delta_E}}
	\end{bmatrix}
	\begin{bmatrix}
		1\\
		\alpha\\
		\frac{\dot{\alpha} \bar{c}}{2 V_{P_1}}\\
		\frac{u}{V_{P_1}}\\
		\frac{q \bar{c}}{2 V_{P_1}}\\
		\delta_E
	\end{bmatrix}
\end{aligned}.
\end{equation}
Furthermore, aerodynamic forces need a transfer from stability to the body frames of the angle of attack $\alpha$ around $y_S$ axis.
\begin{equation}\label{Eq:6}
\begin{aligned}	
	\begin{bmatrix}
		f_{a_x} \\
		f_{a_y}\\
		f_{a_z}
	\end{bmatrix}^B=\begin{bmatrix}
		cos\alpha & 0 & -sin\alpha \\
		0 & 1 & 0 \\
		sin\alpha & 0 & cos\alpha
	\end{bmatrix}^{BS} \begin{bmatrix}
		-D \\
		0 \\
		-L
	\end{bmatrix}^S.
\end{aligned}
\end{equation}
Also, the gravitational acceleration vector $\bm{g}^B$ in the body frame is as follows:
\begin{equation}\label{Eq:6_1}
\begin{bmatrix}
	g_x\\
	g_y\\
	g_z
\end{bmatrix}^B =
\begin{Bmatrix}
	-g\sin(\theta)\\
	g\cos(\theta)\sin(\phi)\\
	g\cos(\theta)\cos(\phi)
\end{Bmatrix}.
\end{equation}
Additionally, rotational kinematic equations are required for transfer from body to inertial frames.
\begin{equation}\label{Eq:7}
\begin{bmatrix}
	\dot{\phi} \\ \dot{\theta} \\ \dot{\psi}
\end{bmatrix}=\begin{bmatrix} 1 & \sin \varphi \tan \theta & \cos \varphi \tan \theta \\ 0 & \cos \varphi & -\sin \varphi \\ 0 & \sin \varphi / \cos \theta & \cos \varphi / \cos \theta \end{bmatrix}\begin{bmatrix} p \\ q \\ r
\end{bmatrix}^B.
\end{equation}
So, the translational kinematic equations using (\ref{Eq:1}), and (\ref{Eq:7}), in the inertial frame is achievable. 
\begin{equation}\label{Eq:8}
	\begin{bmatrix} \dot{x} \\ \dot{y} \\ \dot{z} \end{bmatrix}^E=\begin{bmatrix} \cos \psi \cos \theta & \cos \psi \sin \theta \sin \varphi-\sin \psi \cos \varphi & \cos \psi \sin \theta \cos \varphi+\sin \psi \sin \varphi \\ \sin \psi \cos \theta & \sin \psi \sin \theta \sin \varphi+\cos \psi \cos \varphi & \sin \psi \sin \theta \cos \varphi-\cos \psi \sin \varphi \\ -\sin \theta & \cos \theta \sin \varphi & \cos \theta \cos \varphi \end{bmatrix}\begin{bmatrix} u \\ v \\ w \end{bmatrix}^B.
\end{equation}
Based on Computational Fluid Dynamics (CFD), stability and control derivatives for the Chaka-50 are presented in \cite{zavaree2021modern}. Table (\ref{tab:table1}) summarises these derivatives for two flying scenarios.
Trim conditions in a wings-level flight are calculated for simulation verification utilizing trim equations in \cite{roskam1998airplane}.
The drag equation takes into account the absolute values of $\delta_E$, $i_{H_1}$, and $\alpha_1$.
In addition, the flight path angle $\gamma_1$, motor installation angle $\phi_T$, and horizontal tail incidence angle $i_H$ are all equal to zero.
By solving the trim equation, the elevator deflection $\delta_E$ and required thrust $f_{a_x}$ for a trim flight are obtained and shown in table (\ref{tab:table2}).
The numbers in the aforementioned table (\ref{tab:table2}) are crucial for the validation of the 6-DoF simulation.
\begin{table}[h!]
\begin{center}
	\caption{Stability and control derivatives of Chaka 50 MRJ (1/rad)}
	\label{tab:table1}
	\begin{tabular}{cccccc} 
		\hline
		\begin{tabular}{c}Longitudinal\\Derivatives
		\end{tabular}   & Ideal & \begin{tabular}{c}
		Random \\ Uncertainty \\ ([-10 10])\%
	\end{tabular} & \begin{tabular}{c}Longitudinal\\Derivatives
	\end{tabular}   & Ideal & \begin{tabular}{c}
	Random \\ Uncertainty \\ ([-10 10])\%
\end{tabular}\\
		\hline
		$c_{D_0}$  & 0.0338 & 0.0358 & $c_{L_u}$ & 0.081 & 0.076\\
		$c_{L_0}$   & 0.3180 & 0.3363 & $c_{m_u}$ & -0.039 & -0.041\\
		$c_{m_0}$  & -0.06 & -0.061 & $c_{L_q}$ & 12.53 & 12.56\\
		$c_{D_\alpha}$  & 0.8930 & 0.893 & $c_{m_q}$ & -40.69 & -37.27\\
		$c_{L_\alpha}$  & 14.88 & 14.52 & $c_{D_{\delta_E}}$ & 0.1570 & 01483\\
		$c_{m_\alpha}$  & -11.84 & -11.84 & $c_{L_{\delta_E}}$ & 0.78  & 0.74\\
		$c_{D_u}$ & 0.041 & 0.373 & $c_{m_{\delta_E}}$ & -5.98 & -5.93\\
		\hline
	\end{tabular}
\end{center}
\end{table}
\begin{table}
\begin{center}
	\caption{Trim parameters of Chaka 50 MRJ}
	\label{tab:table2}
	\newcolumntype{C}{>{\centering\arraybackslash}X}
	\begin{tabular}{cc} 
		\hline
		Parameter & Value \\
		\hline
		Required Thrust ($f_{a_x}$)  & 21433.02 (lbs) \\ Required Elevator ($\delta_E$) &  0.39 (deg) \\
		Angle of Attack ($\alpha$) & -2.28 (deg)\\
		\hline
	\end{tabular}
\end{center}
\end{table}
\begin{table}
\begin{center}
	\caption{Chaka 50 MRJ specifics for simulation}
	\label{tab:table3}
	\newcolumntype{C}{>{\centering\arraybackslash}X}
	\begin{tabular}{cccc}
		\hline
		Parameter  & Value & Parameter & Value \\
		\hline
		Wing Area($m^2$) & 43.42 & $I_{xx}$($kg.m^2$) & 378056.535 \\
			Mean Aerodynamic Chord($m$)
		 & 1.216 & $I_{yy}$($kg.m^2$) & 4914073.496 \\
		Span($m$) & 28 & $I_{zz}$($kg.m^2$) & 5670084.803 \\
		Mass($kg$) & 18418.27 & $I_{xz}$($kg.m^2$) & 0 \\
		Initial Speed $V_{P_1}(\frac{m}{s})$ & 160 & Initial height $h_1 (m)$ & 100 \\
		\hline
	\end{tabular}
\end{center}
\end{table}
\subsection{Six-DoF Aircraft Dynamic Model considering Atmospheric Disturbances}
Aircraft landing quality is affected by atmospheric disturbance which is air turbulence in the small areas of the atmosphere that often happens close to the ground. The coordinates of disturbances that result in a loss of lift and altitude are the most hazardous since the aircraft is getting close to the final approach. The atmospheric disturbance is described in the literature as a stochastic process that is characterized by velocity spectra. There are two widely used models that are typically used in flight dynamic simulations:\newline
(1) Dryden Continuous Turbulence Model; and \newline
(2) Von Karman Continuous Turbulence Model \newline
The Dryden atmospheric turbulence is used in this study for two reasons. First, it allows for easier mathematical modeling and covers both linear and rotational components of disturbance velocity.
\begin{equation}\label{Eq:9}
	\begin{aligned}
		&G_u(s) = \sigma_u\sqrt{\frac{2L_u}{\pi 	u_1}}\Bigg[\frac{1}{1+(\frac{L_u}{u_1}s)}\Bigg], \\
		&G_v(s) = \sigma_v\sqrt{\frac{L_v}{\pi u_1}}\Bigg[\frac{1+2\sqrt{3}\frac{L_v}{u_1}s}{(1+\frac{2L_v}{u_1}s)^2}\Bigg], \\
		&G_w(s) = \sigma_w\sqrt{\frac{2L_w}{\pi u_1}}\Bigg[\frac{1+2\sqrt{3}\frac{L_w}{u_1}s}{(1+\frac{2L_w}{u_1}s)^2}\Bigg],
	\end{aligned}
\end{equation}
where according to \cite{mil19808785c}, $L_w$, $L_v$, and $L_u$ are the scaling length;
\begin{equation}\label{Eq:10}
	\begin{aligned}
		&L_u = L_v = \frac{z}{(0.177 + 0.000823z)^{1.2}}, \\
		&L_w = z,
	\end{aligned}
\end{equation}
and $\sigma_u$, $\sigma_v$, and $\sigma_w$ are the intensity of turbulence.
\begin{equation}\label{Eq:11}
	\begin{aligned}
		&\sigma_u = \sigma_v = \frac{\sigma_w}{(0.177 + 0.000823z)^{0.4}}, \\
		&\sigma_w = 0.1u_{20},
	\end{aligned}
\end{equation}
where wind speed at 20 feet is specified by $u_{20}$. The motion equations have now been modified to include the effects of the wind based on \cite{frost1984wind}. In general, the inertial frame is used to compute the wind and its derivatives. But complex computations are required for its transition into the body. As an alternative, one may use the derivatives in the body reference to reach;
where the $\bm{W}^B$ is the wind velocity vector in body frame; $[W]^B = [W_x\ \ W_y\ \ W_z]^T$.
\begin{equation}\label{Eq:12}
	\begin{aligned}
		\dot{W}_x &=\bigg[\frac{\partial W_x}{\partial x}\bigg]^B(u+W_x)+\bigg[\frac{\partial W_x}{\partial y}\bigg]^B(v+W_y)+\bigg[\frac{\partial W_x}{\partial z}\bigg]^B(w+W_z)+\bigg[\frac{\partial W_x}{\partial t}\bigg]^B, \\
		\dot{W}_y &=\bigg[\frac{\partial W_y}{\partial x}\bigg]^B(u+W_x)+\bigg[\frac{\partial W_y}{\partial y}\bigg]^B(v+W_y)+\bigg[\frac{\partial W_y}{\partial z}\bigg]^B(w+W_z)+\bigg[\frac{\partial W_y}{\partial t}\bigg]^B, \\
		\dot{W}_z &=\bigg[\frac{\partial W_z}{\partial x}\bigg]^B(u+W_x)+\bigg[\frac{\partial W_z}{\partial y}\bigg]^B(v+W_y)+\bigg[\frac{\partial W_z}{\partial z}\bigg]^B(w+W_z)+\bigg[\frac{\partial W_z}{\partial t}\bigg]^B.
	\end{aligned}
\end{equation}
The spatial derivatives of the wind speed, which are often stated in the inertial frame, must be transferred to the body frames of reference in (\ref{Eq:12});
\begin{equation}\label{Eq:13}
	[\nabla W]^B = [T]^{BE} [\nabla W]^E [\bar{T}]^{BE}.
\end{equation}
The effect of wind on angular rates $\bm\omega_w^E$ can be defined as a rigid solid air caused by fluid stresses, and is expressed in the inertial frame as;
\begin{equation}\label{Eq:14}
	\begin{aligned}
		[\omega_w]^E &=\frac{1}{2}\bigg[(\frac{\partial W_z}{\partial y}-\frac{\partial W_y}{\partial z})\bigg]^E {i}+ \frac{1}{2}\bigg[(\frac{\partial W_x}{\partial z}-\frac{\partial W_z}{\partial x})\bigg]^E {j}+ \frac{1}{2}\bigg[(\frac{\partial W_y}{\partial x}-\frac{\partial W_x}{\partial y})\bigg]^E {k}.
	\end{aligned}
\end{equation}
The above equation must be transferred to the body axis so as to use in the 6-DoF equation;
\begin{equation}\label{Eq:15}
	\begin{bmatrix}
		p \\
		q \\
		r
	\end{bmatrix}^B = \begin{bmatrix}
	p \\
	q \\
	r
	\end{bmatrix}^B-[T]^{BE} \begin{bmatrix}
		(\frac{\partial W_z}{\partial y}-\frac{\partial W_y}{\partial z}) \\
		(\frac{\partial W_x}{\partial z}-\frac{\partial W_z}{\partial x}) \\
		(\frac{\partial W_y}{\partial x}-\frac{\partial W_x}{\partial y})
	\end{bmatrix}^E
.
\end{equation}
\subsection{Actuator Fault}
A type of failure that affects the plant inputs is an actuator fault. Actuator faults in the aircraft might result from improper maintenance procedures, the age of the material, or improper operation. In this study, the actuator fault is expressed in two terms: the first, is a multiplicative term, which is the elevator's inability to achieve the required amount, and the second, is an additive term, which is the output quantity bias.
\begin{equation}\label{Eq:16}
	\begin{aligned}
		\delta_{E_t} =\ & 0.3\delta_{E_t}-0.7^\circ,\qquad \textrm{If} \ t>12s, \\
		& 0.4\delta_{E_t}+0.6^\circ,\qquad \textrm{If} \ 12s>t>8s, \\
		& 0.5\delta_{E_t}-0.5^\circ,\qquad \textrm{If} \ 8s>t>4s. \\
	\end{aligned}	
\end{equation}
According to equation (\ref{Eq:16}), the elevator operates with $30\%$ power and concurrently biases its output by $-0.7^\circ$ after 12 seconds of flight because it is anticipated that the issue would worsen over time. After 4 and 8 seconds of flight, identical faults with different parameter values will have occurred. In this study, $50\%$ deficiency and $60^\circ$ additive bias after 4 seconds and $30\%$ deficiency and $0.6^\circ$ additive bias after 8 seconds were taken into consideration.

\begin{figure}[!ht]
	\begin{center}
		\includegraphics[width=10.5cm]{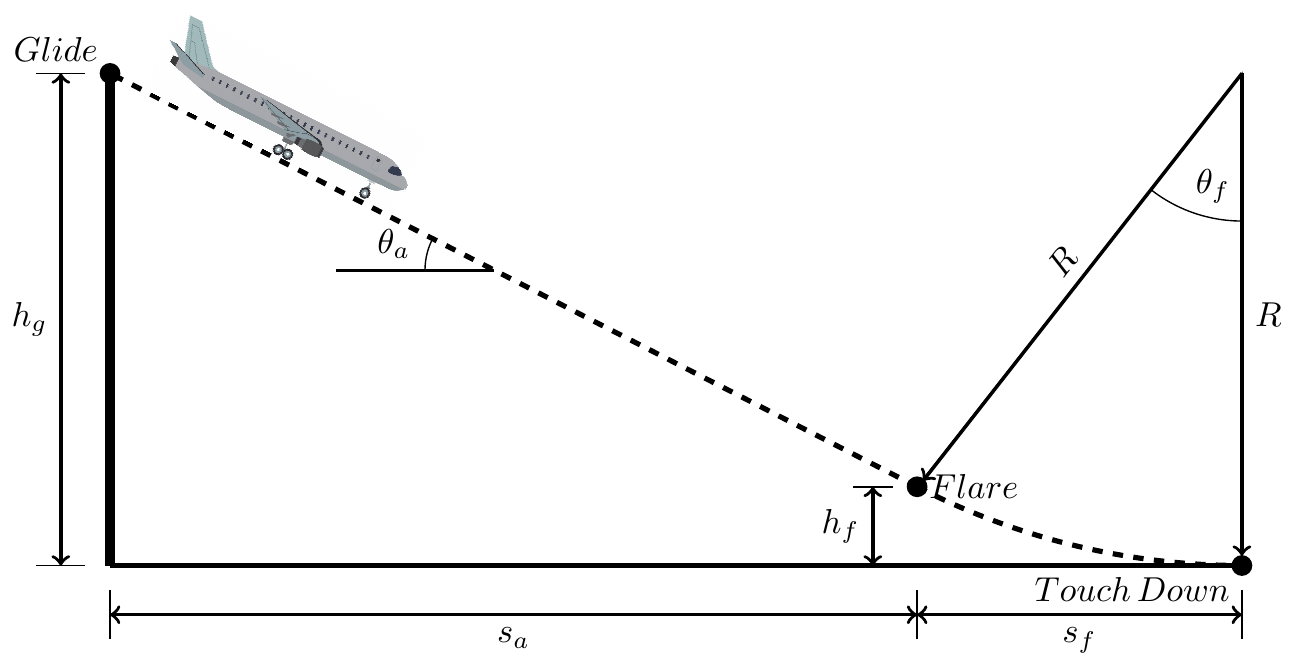}    
		\caption{Landing Path and Landing Distance Diagram}
		\label{fig:3}
	\end{center}
\end{figure}

\section{Landing Path Planning}
Examining Figure \ref{fig:3}, the desired approach angle is $\theta_a \leqslant 3$ according to \cite{book:172651}. In this case assuming aircraft speed in touch down point $V_{TD} = 1.15V_{stall}$, and also speed in flare zone $V_f = 1.23V_{stall}$. So, the formula of flare circular arc is:
\begin{equation}\label{Eq:17}
	R = \frac{V_f^2}{0.2g}.
\end{equation}
Now, in order to calculate $\theta_f$ in each time-step, flare altitude $h_f$ is required.
\begin{equation}\label{Eq:18}
	h_f = R-R\cos\theta_a.
\end{equation}
By considering aforementioned formulas, the approach distance $s_a$, and the flare distance $s_f$ are as follows:
\begin{equation}\label{Eq:19}
	s_a = \frac{-(50 - h_f)}{\tan\theta_a},
\end{equation}
\begin{equation}\label{Eq:20}
	s_f = -R\sin\theta_a.
\end{equation}
Also, by using the formula $s_{td} = s_a+s_f$ to represent touch-down distance, the desired $\theta_f$ at each time step after covering approach distance is produced by:
\begin{equation}\label{Eq:21}
	\theta_f = \arcsin(\frac{x - x_{td}}{R}).
\end{equation}
\section{Dynamic Inversion Auto Landing Structure}
Consider the following formulation to represent the nonlinear dynamic system:
\begin{equation}\label{Eq:22}
	\begin{aligned}
	&\bm{\dot x} = \bm{f}(\bm{x}) + \bm{g}(\bm{x})\bm{u},\ \bm{x}(0) = \bm{x}_0, \\
	&\bm{y} = \bm{h}(\bm{x}),
	\end{aligned}
\end{equation}
where $\bm{x} \in \bm{R}$, $\bm{y} \in \bm{R}$, $\bm{u} \in \bm{R}$ are the vector of state, measurement output, and control input. The goal of the problem is often to design a suitable $\bm{u}$ such that $\bm{y}$ tracks desired $\bm{y}_{des}$. By differentiating the output vector $\bm{y}$ with respect to the state vector $\bm{x}$, this would be accomplished.
\begin{equation}\label{Eq:23}
	\begin{aligned}
		\bm{\dot y} = \frac{\partial \bm{h}}{\partial \bm{x}} \frac{d\bm{x}}{dt} = \frac{\partial \bm{h}}{\partial \bm{x}}\bm{f}(\bm{x})+ \frac{\partial \bm{h}}{\partial \bm{x}}\bm{g}(\bm{x})\bm{u}.
	\end{aligned}
\end{equation}
The error between $\bm{y}$ and $\bm{y}_{des}$ must be eliminated by establishing a first-order dynamic error, which is defined as $\bm{e}_y = \bm{y} - \bm{y}_{des}$:
\begin{equation}\label{Eq:24}
	\begin{aligned}
		&\bm{\dot{e}}_y + \bm{K} \bm{e}_y = 0, \\
		&\bm{\dot{y}} - \bm{\dot{y}}_{des} + \bm{K}(\bm{y} - \bm{y}_{des}) =0.
	\end{aligned}
\end{equation}
The error can be exponentially reduced by calculating the preceding equation while assuming that $\bm{K}$ is a positive definite matrix. Let assume $\bm{F_y}(\bm{x}) = \frac{\partial \bm{h}}{\partial \bm{x}} \bm{f}(\bm{x})$, $\bm{G_y}(\bm{x}) = \frac{\partial \bm{h}}{\partial \bm{x}} \bm{g}(\bm{x})$ and $\bm{g}(\bm{x})$ as an invertible matrix to derive the control signal as follows:
\begin{equation}\label{Eq:25}
	\begin{aligned}
		\bm{u} = [\bm{G}(\bm{x})]^{-1}[\bm{\dot{y}}-\bm{F_y}(\bm{x})].
	\end{aligned}
\end{equation}
By substituting $\bm{\dot{y}}$ as a function of $\bm{e}$:
\begin{equation}\label{Eq:26}
	\begin{aligned}
		\bm{u} = [\bm{G_y}(\bm{x})]^{-1} [\bm{\dot{y}}_{des}-\bm{K}(\bm{y}-\bm{y}_{des})-\bm{F_y}(\bm{x})].
	\end{aligned}
\end{equation}
The main challenge is assigning $\bm{K}$ for error reduction. Extreme non-linearity, couplings, modeling uncertainty, and aerodynamic uncertainty combined with atmospheric disturbances lead to adaptive $\bm{K}$ designation in this problem, which increases complexity. In this section,  the Dynamic Inversion controller was developed based on \cite{AMBATI2017218}. According to aforesaid theories, first-order error in outer loop is considered $\dot{e}_h + k_h e_h = 0$, where $e_h = h - h_{des}$. So the desired path angle is expressed as:
\begin{equation}\label{Eq:27}
	\begin{aligned}
		\theta_{des} = \arcsin \Biggl(\frac{\dot h_{des}-k_h(h - h_{des})}{\sqrt{a_h^2+b_h^2}}\Biggl)-\arctan \biggl (\frac{b_h}{a_h}\biggl),
	\end{aligned}
\end{equation}
where $a_h = u$, and $b_h = v\sin \phi + w \cos \phi$. Then for inner loop control, the first-order error is used respectively. So; 
\begin{equation}\label{Eq:28}
	\begin{aligned}
		\dot\theta - \dot \theta_{des} + k_{\theta}(\theta-\theta_{des})=0.
	\end{aligned}
\end{equation}
Obviously, forces and moments are applied to aircraft in body axis.
 \begin{equation}\label{Eq:29}
 	\begin{aligned}
 		q_{des} = \sec \phi(\dot \theta_{des}-k_{\theta}(\theta - \theta_{des})+r\sin \phi).
 	\end{aligned}
 \end{equation}
Now the control of $q$ in order to track $q_{des}$ is applicable. So the first-order error is applied:
 \begin{equation}\label{Eq:30}
	\begin{aligned}
		\dot{q} - \dot{q}_{des} + k_q(q - q_{des})=0.
	\end{aligned}
\end{equation}
By driving $\dot{q}$ in equation (\ref{Eq:1}) and substituting equation (\ref{Eq:30}), the desired longitudinal control input $\delta_{E}$ is accessible.
 \begin{equation}\label{Eq:31}
	\begin{aligned}
		\delta_{E} = \frac{I_y(\dot{q}_{des} - k_q(q - q_{des}))+(I_x-I_z)rp-M_A-M_{\delta_{E}}\delta_{E}}{M_{\delta_{E}}},
	\end{aligned}
\end{equation}
where $M_{\delta_E} = \bar{q}S\bar{c}c_{m_{\delta_E}}$. \newline
\textbf{Remark:} It is apparent that the elevator deflection computation needs to desired pitch angle and desired pitch rate states. But in the FQL method, it will be seen later that the control signals are computed by just the desired pitch angle state during the trajectory tracking phase.
\section{Fuzzy Q-learning Auto-landing Structure}
Because of their narrow Mean Aerodynamic Chord (MAC), TBW aircraft typically exhibit insufficient longitudinal stability.
More specifically, this finding may be confirmed by evaluating the Phugoid and Short-period modes of the Boeing N+3 TTBW \cite{nguyen2022dynamic} and Chaka 50 against the Cessna 172 \cite{cetin2018}.
Table (\ref{tab:table4}) contains an overview of the numerical data for the longitudinal modes of the aforementioned transport aircraft. This table can support the claim made above. To be clear, due to their superior manoeuvrability, the Chaka and Boeing, which have similar designs, experience poor longitudinal stability qualities. The Cessna 172, on the other hand, benefits from stable dynamics behaviour.
For clarification, the damping ratio of the Chaka 50's Phugoid mode causes low-stability behavior resulting in long-term oscillations as depicted in \cite{act11120374}.
\begin{table}[h!]
\begin{center}
	\caption{Longitudinal Dynamics Characteristic}
	\label{tab:table4}
	\begin{tabular}{ccc} 
		\hline
		Aircraft\ Roots& Short Period Roots & Phugoid Roots \\
		\hline
		Chaka 50 & $-0.8 \pm 0.61i$ & $-0.0064 \pm 0.05i$ \\
		Cessna 172 & $-3.23 \pm 5.71i$ & $-0.025 \pm 0.19i$  \\
		Boeing N+3 & $-0.35 \pm 0.35i$ & $-0.0082 \pm 0.07i$  \\
		\hline
	\end{tabular}
\end{center}
\end{table}
\subsection{MDP Definition in Auto-landing}
A sequential decision-making process, such as an auto-landing problem, must be formalized as MDPs since one action affects not only the next state and its immediate reward but also forthcoming states and their future rewards \cite{sutton2018reinforcement}. To be clear, at each time-step $t$, the controller receives an state observation from the 6-DoF simulation, including $\theta_t \in \bm{S_1}$ and $\dot{\theta_t} \in \bm{S_2}$. Based on it, the controller specifies an action, $\delta_{E_t} \in \bm{A}(\bm{s})$, which is the elevator deflection. The simulation runs, and in the following time step $t+1$, the controller receives a reward $R_{t+1} \in \bm{R}$ to assess its performance and find itself in the next state $\theta_{t+1}, \dot\theta_{t+1}$ until reaching to terminal state $\theta_{T}, \dot\theta_{T}$.
\begin{equation}\label{Eq:32}
\theta_0,\ \dot{\theta}_0,\ \delta_{E_0},\ R_1,\ \theta_1,\ \dot{\theta}_1,\ \delta_{E_1},\ R_2,\ ... \ \theta_T,\ \dot{\theta}_T.
\end{equation}
Here, $\theta_{0}$, and $\dot{\theta}_0$ are the initial states, and $\delta_{E_0}$ is the initial elevator deflection, moreover $R_t$ defines the instant reward at time-step $t$. A random pick of $R_t$, $\theta_t$, and $\dot\theta_t$ have a clear discrete probability distribution that is solely reliant on the past state-action in this issue. Thus, by treating $\theta_t$ and $\dot\theta_t$ as states, the Markov property is satisfied because they contain complete information about all aspects of the controller-aircraft (agent-environment) interaction history that matter in the future. 
So, for all $\theta_c$, $\theta_p \in \bm{S}_1$, $\dot{\theta}_c$, $\dot{\theta}_p \in \bm{S}_2$, and $\delta_E \in \bm{A}(\bm{s})$, the equation of the MDP problem is defined as $P$;
\begin{equation}\label{Eq:33}
	\begin{aligned}
		 P(\theta_c, {\dot{\theta}}_c\ |\ \theta_p, \dot{\theta}_p, \delta_E) = Pr \{\theta_{t} = \theta_c, \dot{\theta_t} = {\dot{\theta}}_c\ |\ \theta_{t-1} = \theta_p, \dot\theta_{t-1} = {\dot{\theta}}_p, \delta_{E_t} = \delta_E\},
	\end{aligned}
\end{equation}
where $c$ and $p$ are used for the current state and the previous state, respectively. In this problem, according to table (\ref{tab:table5}), equation (\ref{Eq:33}) always has a deterministic numerical amount in $[0\  1]$ for all states. 
All states are well-defined and each receives a distinct reward.
\begin{equation}\label{Eq:34}
	\begin{aligned}
		{\sum}_{\theta_c \in \bm{S}_1}{\sum}_{\dot\theta_c \in \bm{S}_2}
		P(\theta_c, {\dot{\theta}}_c \ |\ \theta_p, \dot{\theta}_p, \delta_E) =1.		
	\end{aligned}
\end{equation}
The goal of finite MDP is to design a policy that maximizes reward over time. 
To find an optimal policy for taking $\delta_E$ in state $\theta$, $\dot{\theta}$, the state-action value function $\bm{Q_}\pi(\theta, \dot\theta, \delta_{E})$ must be maximized, which is defined as the expected return as the sum of discounted instant rewards by starting from one specific state and pursuing policy $\pi$ to terminal state $\theta_T$, $\dot\theta_T$:
\begin{equation}\label{Eq:35}
\begin{aligned}
	\bm{Q_}\pi(\theta,\,\dot\theta,\,\delta_E) =  \mathbb{E}_\pi \bigg[ {\sum}_{k=0}^{\infty}\gamma^k R_{t+k+1}\ \bigg|\  \theta_{t} = \theta,\, \dot{\theta}_t = \dot{\theta},\, \delta_{E_t}=\delta_E \bigg],
\end{aligned}
\end{equation}
where $0<\gamma<1$ is the discount factor, whereas $\gamma \simeq 0$ denotes the agent nearsightedness, and $\gamma \simeq 1$ means the agent is long-sighted.

\subsection{Structure of Fuzzy Q-learning Controller}
\begin{figure}
	\begin{center}
		\includegraphics[width=12.5cm]{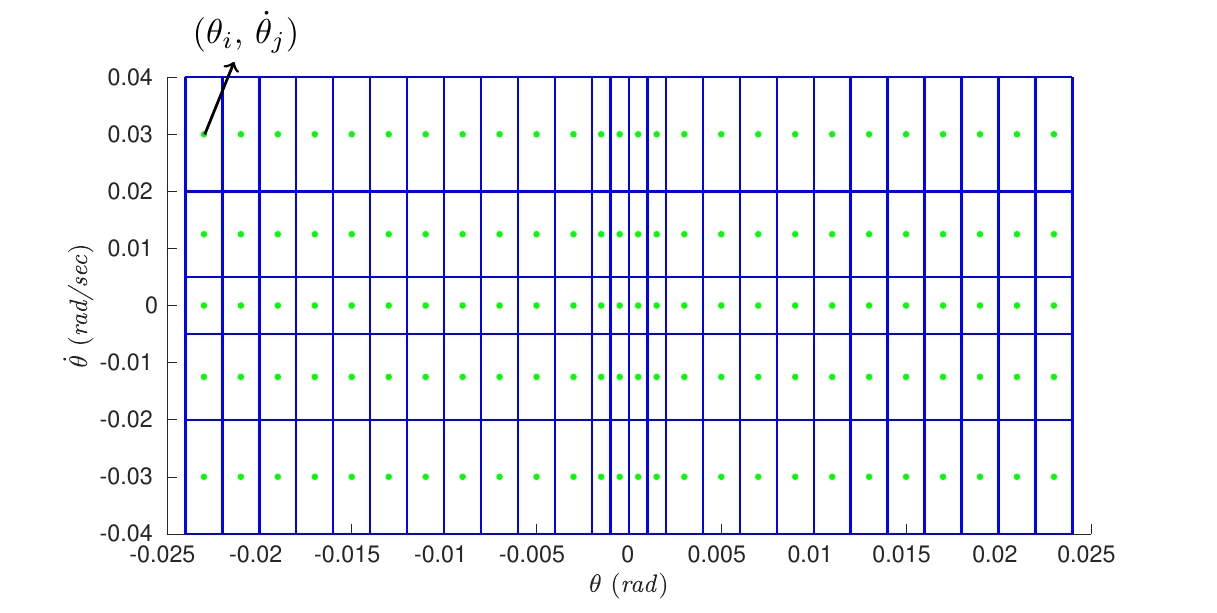}
		\caption{Grid of state variables used for tabular Q-learning (The center of each cell, which is used to compute the membership function of the cell is shown by a circle point.)}
		\label{fig:4}
	\end{center}
\end{figure}

In the current study, Q-learning, an early breakthrough in Reinforcement Learning, is used to directly approximate the optimal elevator selection policy in each condition \cite{watkins1992q}. Q-Learning is an off-policy, model-free control algorithm based on the Temporal Difference (TD) method. 
In general, the system state ($\theta_{t},\dot{\theta}_t$) is obtained each time step, and by using the current elevator command ($\delta_{E_t}$) the action selection policy will be updated. 
Because of the highly nonlinear and poor stability characteristics of TBW aircraft, the Q-learning implementation results can be unsatisfactory   without employing continuous state-action \cite{act11120374}. 
Accordingly, two Q-tables are trained by the Fuzzy and basic Q-learning methodologies. Then, they are Incorporated in different auto-landing scenarios through a novel technique namely the Fuzzy Action Assignment (FAA), which can be introduced as a general connector between various trained Q-tables and continuous environments. Instead of computing a discrete greedy action in a particular state $\theta$, $\dot\theta$, FAA technique assigns a relative weight (also defined as the validity function or membership function) to each cell of the grid of system states (see Figure \ref{fig:4}). This assignment is performed based on the current value of the state-action value function.
The membership function of each grid cell with centers $\theta_i$ and $\dot\theta_j$ is described as follows:
\begin{equation}\label{Eq:36}
	\begin{aligned}
		& MF_{i,j} = \exp\left(-\frac{1}{2}\left(\frac{\theta_t-\theta_{i_t}}{\sigma_{\theta}}\right)^2\right) \exp\left(-\frac{1}{2}\left(\frac{q_t-\dot{\theta}_{j_t}}{\sigma_{\dot{\theta}}}\right)^2\right),
	\end{aligned}
\end{equation}
where $\sigma_{\theta}$, and $\sigma_{\dot{\theta}}$ define the validity widths of the membership functions. 
In this study, the elevator commands of TBW aircraft are specified into $-0.25$ to $+0.25$ radians with $0.025$ intervals, corresponding to 21 elevator deflections. Also, the $\epsilon$-greedy action selection strategy with epsilon decay is utilized in this research so as to select greedy elevator commands in the last episodes (when the trained policy is near-optimum).
\begin{equation}\label{Eq:37}
	\delta_{E_t} =
	\begin{cases}
		{\arg\max}_{\delta_{E}}\  \bm{Q}(\theta_{t}, \dot\theta_{t}, \delta_{E}) & \text{with probability $1-\epsilon$}\\
		\text{random action} & \text{with probability $\epsilon$}\\
	\end{cases}
\end{equation}
Following that, $\delta_E$ is determined using a weighted average of neighbor membership functions at each time step as regards:
\begin{equation}\label{Eq:38}
	\begin{aligned}
		\delta_{E_t}=\frac{\sum_{i}\sum_j MF_{{i,j}_{t}}\ {\arg\max}_{\delta_{E}} \, \bm{Q}(\theta_{i_t}, \dot\theta_{j_t}, \delta_{E})}{\sum_i\sum_j MF_{{i,j}_{t}}}.
	\end{aligned}
\end{equation}
The computed elevator deflection is applied to the aircraft 6-DoF simulation environment and receives a scalar reward signal as performance feedback which is defined in the Reward Function Definition section.

\subsubsection{State-action Value Function Updating Rule}
In this study, the state-action value function directly estimates the optimal Q-table controller which is performed over episodes utilizing Fuzzy rules. The base updating rule equation is unchanged but the calculation of its different parts is improved dramatically.
So in this part, the updating formalization of the terms of updating function is discussed. In this case, $Q_f$ defines as the value of state-action in the previous time step that is computed with its neighbor grids as:
\begin{equation}\label{Eq:42}
	\begin{aligned}
		{Q}_f(\theta_{t}, \dot{\theta_t},  \delta_{E_t})=\frac{\sum_{i}\sum_j MF_{{i,j}_{t}} \, {Q}(\theta_{i_t}, \dot\theta_{j_t}, \delta_{E})}{\sum_i\sum_j MF_{{i,j}_t}}.
	\end{aligned}
\end{equation}
Then, estimating the optimal state-action value of the next time step using a Fuzzy scheme is important in order to, find an estimation of a sequence of best elevator deflection values that will be selected from that state until the terminal state. But note that the quota of neighbor elevators is seen using the membership function. The reason for this computation is obvious. The effect of one specific $\delta_E$ is not only on one pair of $\theta$, $\dot\theta$ but also has a nearly similar effect on their near neighbors. So the Fuzzy optimal future value is as follows;
\begin{equation}\label{Eq:43}
	\begin{aligned}
		\max_{\delta_{E}} {Q_f}(\theta_{t+1}, \dot\theta_{t+1}, \delta_{E})=\frac{\sum_{i}\sum_j MF_{{i,j}_{t+1}}\ {\max}_{\delta_{E}} \, {Q}(\theta_{i_{t+1}}, \dot\theta_{j_{t+1}}, \delta_{E})}{\sum_i\sum_j MF_{{i,j}_{t+1}}},
	\end{aligned}
\end{equation}
where the membership function is related to the next time-step. Another step is to calculate the Temporal Difference (TD) in Fuzzy format. Therefore, the TD formulation is as follows;
\begin{equation}\label{Eq:44}
	TD_{{i,j}_t} = \frac{\sum_{i}\sum_j MF_{{i,j}_{t}}\ \bigg[R_{t+1}+ \gamma \, \underset{\delta_{E}}{max} \, {Q}_f(\theta_{t+1}, \dot\theta_{t+1},  \delta_{E})-{Q}_f(\theta_{t}, \dot{\theta_t},  \delta_{E_t})\bigg]}{\sum_{i}\sum_j MF_t},
\end{equation}
where the membership function in this step is related to the previous time-step.
The pseudocode of the proposed control strategy is summarized in Algorithm \ref{alg:one}.
\subsubsection{Fuzzy Action Assignment Structure}
As discussed earlier, the FAA receives optimal policies produced by various continuous or discrete RL algorithms. Then generates continuous actions for different control problems based on Fuzzy methodologies. Its implication are easy-going using just two equations \ref{Eq:36}, and \ref{Eq:38}. The pseudocode of FAA is explained in algorithm \ref{alg:two}.
\subsubsection{Reward Function Definition}
The definition of an appropriate reward function is critical to the learning processes' convergence.
Therefore, the reward function design and hyper-parameter adjustment have received significant attention in this study likewise.
In this manner, the reward function is generated in three phases and comprises plant states such as $\theta, q, \delta_E$.

To begin with, in order to limit the elevator's high-frequency deflecting, severe penalties in the case of aggressive elevator selection are required. This penalty is performed whenever an elevator changes more than 0.1 radians.
\begin{equation}\label{Eq:39}
	R_t =-10000, \quad \textrm{If} \ \left(|\delta_{E_t}|-|\delta_{E_{t-1}}|\right)>5.73^\circ.
\end{equation}
The reward function will then be calculated as follows if the aircraft is close to the desired angle and the elevator operation frequency is reasonable.
\begin{equation}\label{Eq:40}
	\begin{aligned}
		R_t =\
		& (300, \qquad \textrm{If} \ |e_{\theta_t}|< 0.05^\circ),   \\ + \
		& (300, \qquad \textrm{If} \  |e_{\theta_t}|< 0.02^\circ), \ \\+ \
		& (400, \qquad \textrm{If} \ |q_{t}|< 0.04^\circ), \ \\+ \
		& (600, \qquad \textrm{If} \ |q_{t}|< 0.02^\circ), \ \\+ \
		& (800, \qquad \textrm{If} \ |q_{t}|< 0.005^\circ), \
	\end{aligned}
\end{equation}
where $e_{\theta_t} = \theta_{t} - \theta_{des}$ is the proportional error.
This definition first examines the state of pitch tracking. The controller then detects and prioritizes fewer pitch rates using higher reward allocations in the last episodes. The concepts stated above were defined for learning convergence. In other words, they are activated when the pitch angles are near the desired values. However, it is essential to guide the learning process in early episodes with another phrase. As a result, if none of the above two requirements are satisfied, we should urge the air vehicle to proceed at the desired angle. Using the following reward function, this demand is achievable:
\begin{equation}\label{Eq:41}
	R_t = - (100 \times |e_{\theta_t}|)^2 - (40 \times |q_{t}|)^2.
\end{equation}
As a result, the further the system deviates from the desired state, the lower the reward. In addition, to avoid excessive pitch rates, a derivative term (the second term) has been added into reward function.
More specifically, the presence of the pitch rate ($q_t$) in equation (\ref{Eq:40}), as well as its weight in equation (\ref{Eq:41}), influences the convergence rate significantly.

\begin{algorithm}
	\caption{Fuzzy Q-learning Aircraft Attitude Controller}\label{alg:one}
	\KwData{Learning Rate $\alpha$, Discount Factor $\gamma$, Desired Angle $\theta_{des}=1$deg, Validity Widths $\sigma_{\theta}, \sigma_{\dot{\theta}}$, Elevator Deflections $\delta_{E}$, Pitch (Rate) Angle Intervals $\theta$, $\dot{\theta}$.}
	\KwResult{$\bm{Q}_{\pi^*}(\theta, \dot\theta, \delta_{E})$}
	$\bm{Q}(\theta_0, \dot\theta_{0}, \delta_{E_0}) \gets 0$; \textbf{for all}$\ \theta\in {\bm{S}}_1,\dot{\theta}\in {\bm{S}}_2, \delta_E\in\bm{A}(\bm{s})$\;
	\For{Episode Number = 1 to 20000}{
		Initialize 6-DoF simulation with a random $\theta_0 \in [0\ 2]$ deg. \\
		\For{time-step (0.01) = 0 to 5 sec}{
		\eIf{$\epsilon <$ random number $ \in [0\ 1]$}{
			$\delta_{E_t} \gets random \ \delta_E\in\bm{A}(\bm{s}) $\;
		}{ \For{i,j = 1 to length I,J}{
				$MF_{i,j_t} \gets$ Compute membership function eq \ref{Eq:36}\;
				$\delta_{E_t} \gets$ Compute $\delta_{E_t}$ eq \ref{Eq:38}\;
				}
			}
			Execute 6-DoF simulation using computed  $\delta_{E_t}$, observe $R_{t+1}$, $\theta_{t+1}$, $\dot\theta_{t+1}$\;
			\For{i,j = 1 to length I,J}{
			$MF_{{i,j}_{t+1}} \gets$ Compute membership function eq \ref{Eq:36}\;
			${Q_f}(\theta_t, \dot\theta_t, \delta_{E_t}) \gets {Q}(\theta_t, \dot\theta_t, \delta_{E_t})$ Computeed by eq \ref{Eq:42}\;
			$\max_{\delta_{E}} {Q_f}(\theta_{t+1}, \dot\theta_{t+1}, \delta_{E}) \gets \max_{\delta_{E}} {Q}(\theta_{t+1}, \dot\theta_{t+1}, \delta_{E})$ Computeed by eq \ref{Eq:43}\;
			${Q}(\theta_{i_t}, \dot\theta_{j_t}, \delta_{E_t}) \gets {Q}(\theta_{i_t}, \dot\theta_{j_t}, \delta_{E_t}) + \alpha \bigg[TD_{i,j_{t}} \bigg]$ Based on eq \ref{Eq:44}\;
		}
	Substitute simulation parameters in time-step $t$ with $t+1$.
			}
		}
	\textbf{Return} $\bm{Q_}{\pi^*}(\theta, \dot\theta, \delta_{E})$
\end{algorithm}
\begin{algorithm}
	\caption{Fuzzy Action Assignment Scheme}\label{alg:two}
	\KwData{$\bm{Q_}{\pi^*}(\theta, \dot\theta, \delta_{E})$}
	\KwResult{Contineuous Elevator Deflection $\delta_E$}
	Initialize 6-DoF simulation using predefined initial conditions\;
	\For{time-step (0.01) = 0 to 5 sec}{\For{i,j = 1 to length I,J}{
		$MF_{i,j_t} \gets$ Compute membership function eq \ref{Eq:36}\;
		$\delta_{E_t} \gets$ Compute $\delta_{E_t}$ eq \ref{Eq:38}\;
		}
	Execute 6-DoF simulation using computed $\delta_{E_t}$, compute the next system states\;
	Substitute simulation parameters in time-step $t$ with $t+1$.
	}
\end{algorithm}
\section{Simulation Results and Discussion}
\begin{figure}[h!]
	\begin{center}
		\includegraphics[width=13.5cm]{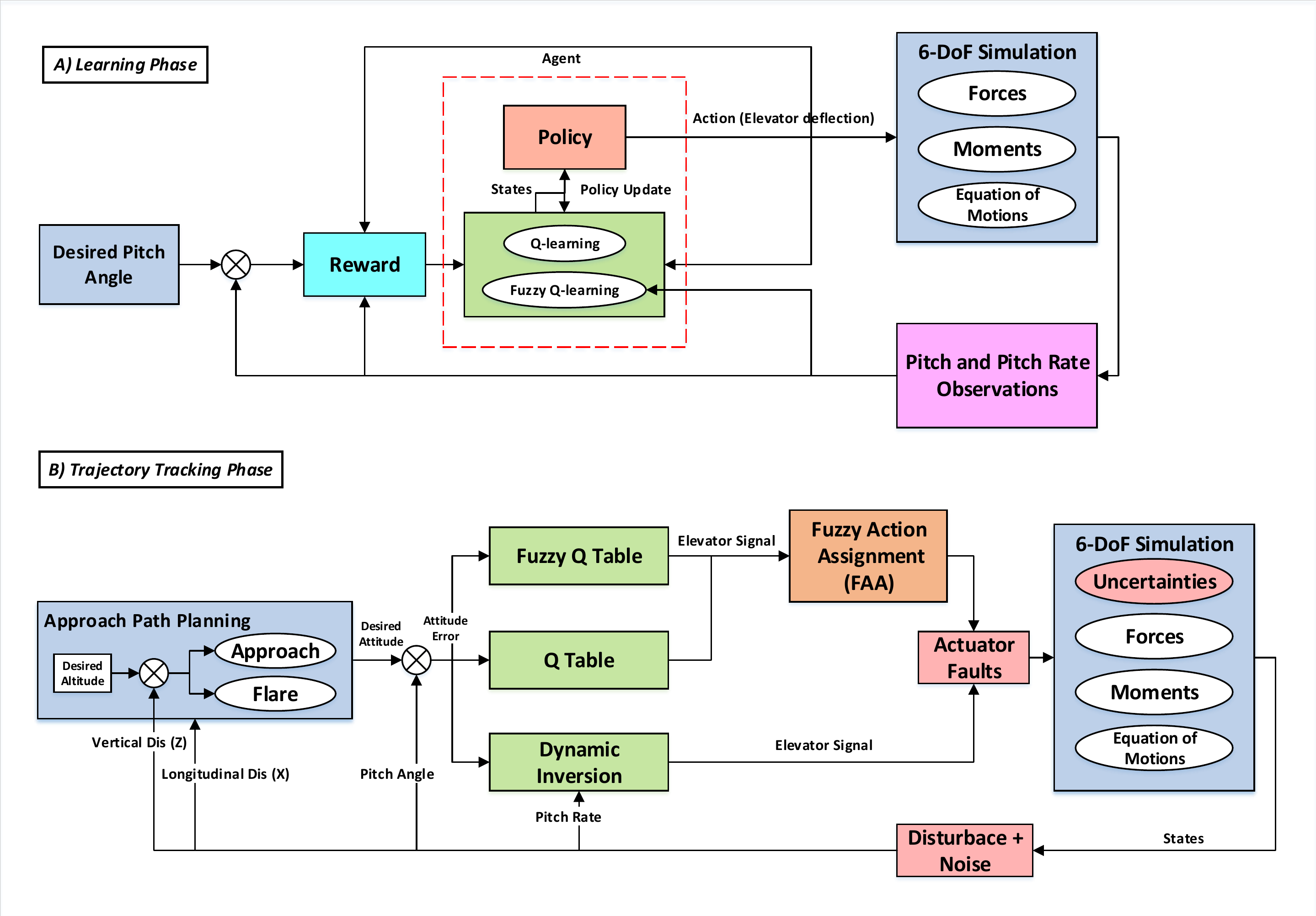}    
		\caption{Block diagram of learning and trajectory tracking phases of all 3 methods}
		\label{fig:5}
	\end{center}
\end{figure}

The auto-landing control of an innovative TBW regional jet aircraft is developed utilizing Fuzzy Q-learning (FQL). The development process and the justification of the findings will therefore be touched upon in this section. The procedure is divided into two segments, as will be detailed later. The first is the learning phase, and the second is the trajectory tracking phase, which is the execution of the optimal policy. The proposed approach is then compared with Dynamic Inversion, a well-known robust control method according to studies. A novel continuous action selection method is developed in this study to function as a useful link between every trained Q-table and environment, as shown in figure \ref{fig:5}.

In the learning phase, the desired pitch angle is set to 1 degree, and the initial pitch angle is selected as a random number between 0 and 2 degrees during episodes. The observation vector contains the pitch angle $\theta$, and pitch rate $(q)$ in each time step. Furthermore, the pitch and pitch angle rate state intervals are defined as a 3D table alongside with action vector (Elevator deflection intervals) which are located in the policy block. This block selects an action based on the $\epsilon$-greedy strategy. The FQL core receives the reward, observation vector, and the relevant states of the pitch angle and pitch rate from the policy block (Figure \ref{fig:5}) and updates the policy based on the aforementioned algorithm \ref{alg:one}. Also, Q-learning receives all mentioned signals except the observation vector for policy updating.

The second phase is specified for the auto-landing control. In this case, the path planning block generates desired pitch angle in each time step. Then, one of the controllers out of three generates elevator signals. The output of the Fuzzy Q-table and Q-table is sent to the FAA block to produce continuous action (elevator deflections) based on algorithm \ref{alg:two}. Several scenarios are defined in this stage including actuator faults, model uncertainties, and atmospheric disturbance plus sensor measurement noise. The parameters required for simulation are gathered in table \ref{tab:table5}. The positive part of pitch angle and pitch rate intervals are not mentioned in this table because of considering similar intervals. Useful to mention that all steps are developed in MATLAB R2022a and the PC is characterized by an 8 cores processor with 2.30 GHz, and 8 GiB RAM.

\begin{figure}[h!]
	\begin{center}
		\includegraphics[width=14.cm]{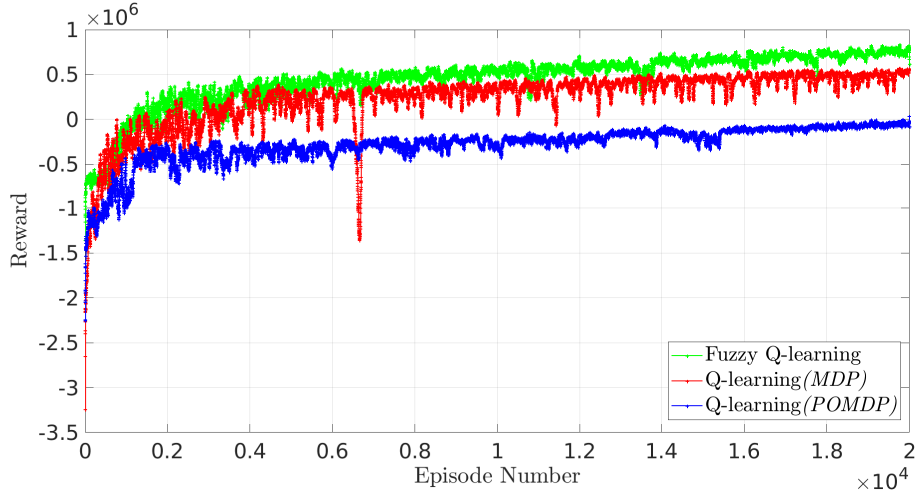}
		\caption{Learning result of Q-learning and Fuzzy Q-learning during $20000$ Episodes}
		\label{fig:6}
	\end{center}
\end{figure}

\begin{figure}
	\begin{center}
		\includegraphics[width=14.5cm]{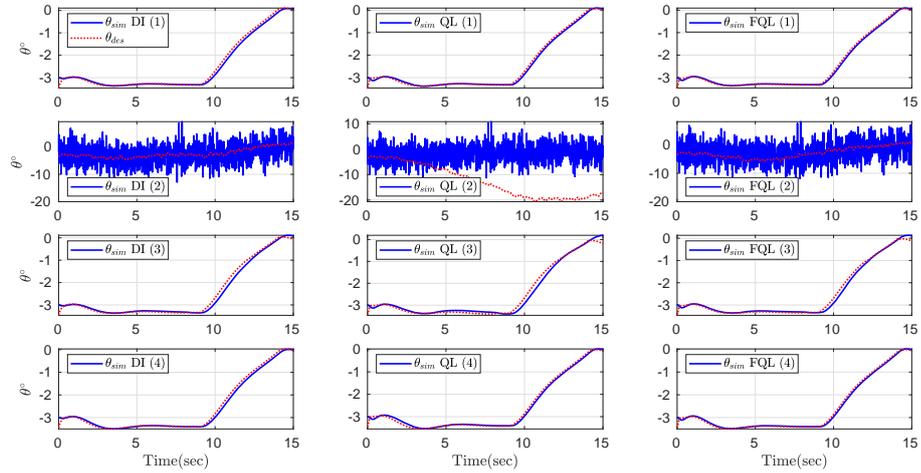}
		\caption{Attitude tracking of all 3 methods in (1) Ideal flight conditions (2) Atmospheric disturbance and sensor measurement noise (3) Actuator faults (4) Model parameters uncertainties.}
		\label{fig:7}
	\end{center}
\end{figure}

The learning results of the two methods are shown in figure \ref{fig:6}. The main difference between MDP and POMDP models is observing the pitch rate in the MDP problem definition which in the POMDP is omitted. Obviously, the FQL conquered in comparison with Q-learning, and this insignificant difference will prove the robustness of the FQL later. Furthermore, its fluctuations are less than Q-learning (MDP). The POMDP policy is not included in the trajectory tracking phase owing to its unsuccessful findings \cite{act11120374}. The results of attitude tracking are gathered in figure \ref{fig:7}. The first row is specified for ideal flight conditions. All three methods were prosperous to track the desired angle. But the main difference is related to attitude tracking errors ($TE_{\theta}$), altitude tracking errors ($TE_{h}$), and control effort ($CE$) which are defined as follows;
\begin{equation}\label{Eq:45}
	TE_{\theta} = \frac{\int_{0}^t |\theta_{t} - \theta_{des}| dt}{t},
\end{equation}
\begin{equation}\label{Eq:46}
	TE_{h} = \frac{\int_{0}^t |h_{t} - h_{des}| dt}{t},
\end{equation}
\begin{equation}\label{Eq:47}
	CE = \frac{\int_{0}^t |\delta_{E_t}| dt}{t}.
\end{equation}
The second row contains simulation results in atmospheric disturbance and sensor measurement noise. The DI and FQL are able to track tough high-frequency responses. Although the path-planning block tried to generate less desired pith angles to guide the aircraft to the desired trajectory, the QL diverged. The third row includes the results of actuator faults and finally, the last row gathers the results of model parameter uncertainties which the performance of all controllers are reasonable.
\begin{table}
\begin{center}
	\caption{Controller parameters in learning and tracking phases}
	\label{tab:table5}
	\begin{tabular}{ccc}
		\hline
		\textbf{Parameter} & \textbf{Definition} &\textbf{Value}  \\
		\hline
		Epsilon($\epsilon$) & Exploration Probability &$[0.1: \textcolor{blue}{3e{-6}}: 0.04]$  \\
		Alpha($\alpha$) & Learning Rate & $[0.02: \textcolor{blue}{9e{-7}}: 0.002]$ \\
		Gamma($\gamma$)& Discount Factor& $0.99$ \\
		Episode number& - &$20000$ \\
		$\theta$(rad) & Pitch Angle Intervals & $[-10, -0.024: \textcolor{blue}{0.002}: -0.002, -0.001, 0]$\\
		$\dot\theta$(rad) & Pitch Rate Intervals & $[-10, -0.04, -0.02, -0.005]$\\
		I & Adjacent pitch grids & $[i-2: i+2]$\\
		J & Adjacent pitch rate grids & $[j-2: j+2]$\\
		$k_h$ & Altitude Control Coefficient & 1.3 \\
		$k_{\theta}$ & Pitch Control Coefficient & 5\\
		$k_q$ & Pitch Rate Control Coefficient & 10\\
		\hline
	\end{tabular}
\end{center}
\end{table}

\begin{figure}
	\begin{center}
		\includegraphics[width=14.5cm]{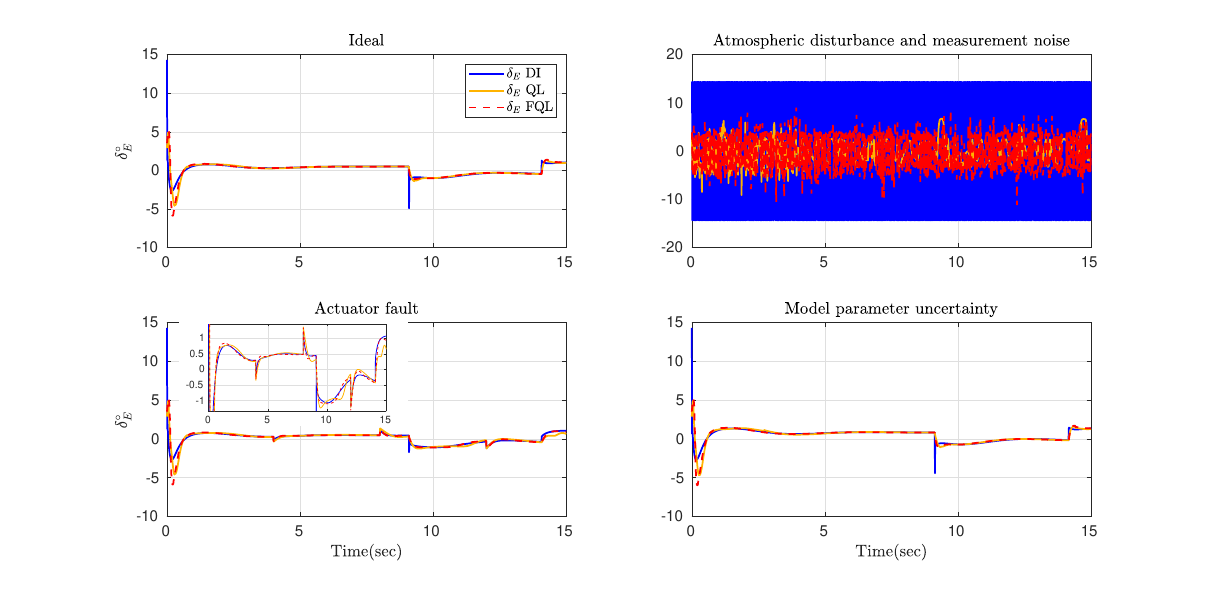}
		\caption{Computed elevator deflections of all 3 methods in different flight conditions}
		\label{fig:8}
	\end{center}
\end{figure}

According to figure \ref{fig:8}, the elevator deflection of all three methods explains significant data. To clarify, the first subplot indicates a larger initial deflection for DI in ideal flying conditions than the others. Furthermore, the second subplot demonstrates the poor performance of DI in the face of sensor noise and atmospheric disturbances. This result may be successful in trajectory tracking but is unable to be applicable. The third subplot is drawn for faulty elevator deflections with conspicuous jumps in all schemes. Finally, in contrast to the first subplot, the last subplot yields roughly analogous findings. It is useful to note that the DI overshoot before 10 seconds demonstrates its sensitivity to the commencing flare phase.

The next figure \ref{fig:9} demonstrates the altitude tracking of all three controllers. In an overview, the first, third, and last subplots prove the robustness of all three methodologies theoretically. But the second subplot is an exception. The QL controller is unable to accomplish its task properly but the FQL and DI have reasonable findings although the performance of DI is not useful for real applications. However, it is noticeable that the altitude tracking error of FQL in the middle of the approach distance is larger than DI. But during the flare phase that is more critical, the story changes, where the FQL surpasses DI.
\begin{figure}
	\begin{center}
		\includegraphics[width=14.5cm]{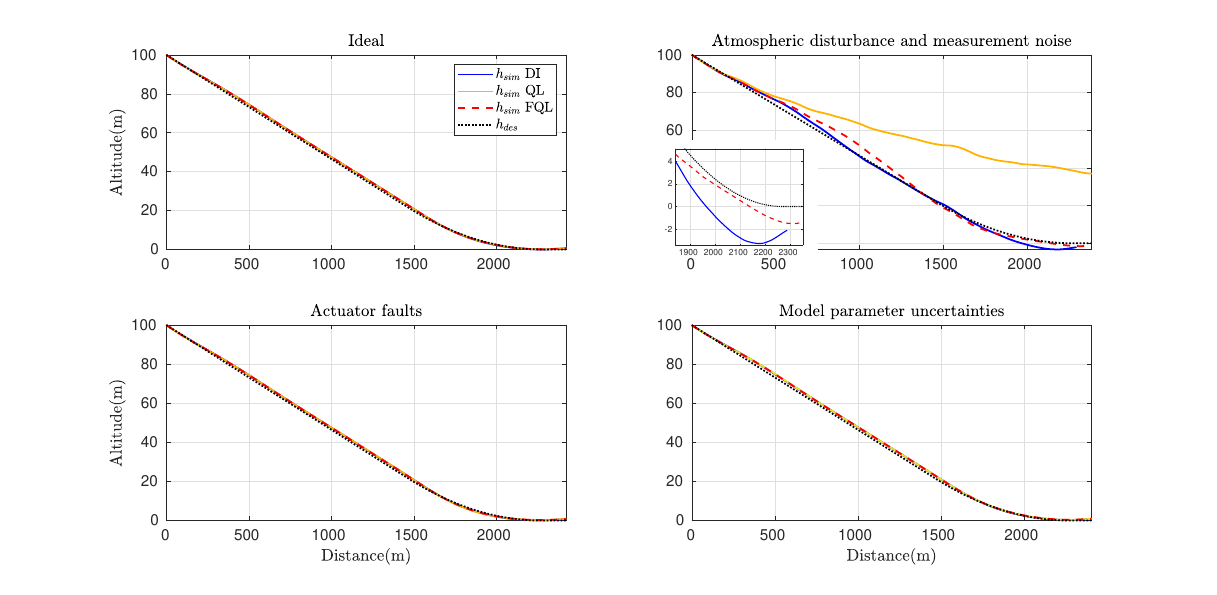}
		\caption{Altitude tracking of all 3 methods in various scenarios}
		\label{fig:9}
	\end{center}
\end{figure}

\begin{figure}[h!]
	\begin{center}
		\includegraphics[width=14.5cm]{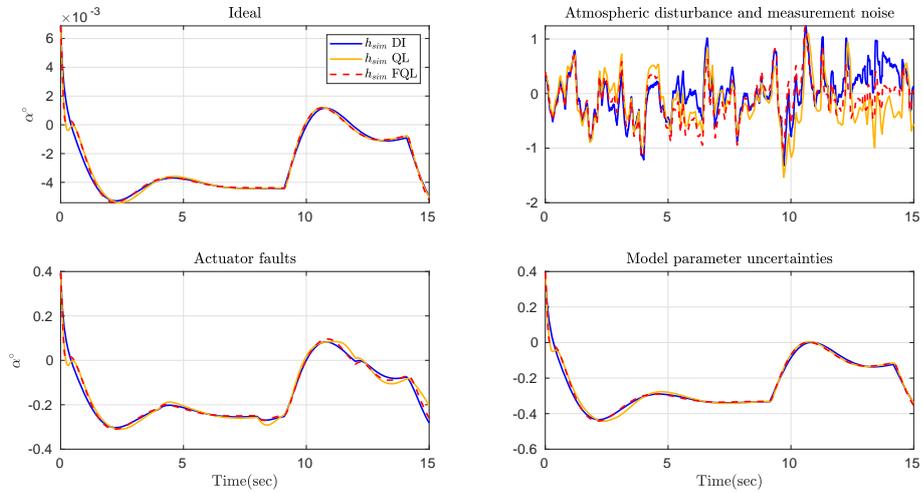}
		\caption{Aircraft Angle of Attack results in different flight simulations}
		\label{fig:10}
	\end{center}
\end{figure}

The angle of attack (AoA) results of 15 seconds flight simulations are shown in figure \ref{fig:10}. The less magnitude of AoA changes is clear in the first subplot compared to others. After the first subplot, the actuator faults have less influence on AoA although, there are some little jumps. The second subplot demonstrates an AoA change between -1 to 1 degree which is more than others despite its noisy specificity. This can amplify the drag where its consequences are clear in the next figure. Generally, the effect of actuator faults on AoA is less than model uncertainties.

The last figure depicts the speed of the airplane along its longitudinal body axis. According to the path planning section, the speed of aircraft at the touch-down point should be $161\frac{m}{sec}$. The simulation of all flight conditions yields satisfactory results, however, the second subplot reveals a problem with DI. The performance of this controller leads the aircraft speed to stall margin, and the authors attribute this occurrence to high drag production caused by the elevator.
\begin{figure}[h!]
	\begin{center}
		\includegraphics[width=14.5cm]{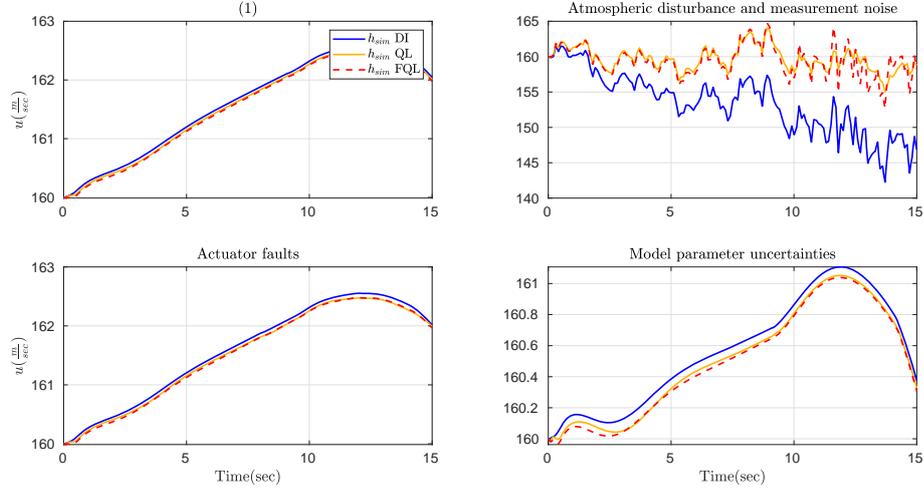}
		\caption{Simulated longitudinal speed results during various scenarios}
		\label{fig:11}
	\end{center}
\end{figure}

A summary of results is gathered numerically in table \ref{tab6} for better comprehension. At a first glance, the superiority of FQL is noticeable in attitude tracking of all flight conditions except in presence of noise and disturbance where DI depicts better findings apparently. But the performance of DI in this circumstance is called into question owing to its control effort. More precisely about ideal conditions, the altitude tracking errors of QL and FQL are slightly better than DI, unlike their control effort. However, the differences are insignificant. As previously stated in the second scenario, despite DI superiority in pitch and altitude tracking errors, elevator deflection is saturated drastically.
In the simulated flight influenced by actuator faults, the FQL controller is the forerunner. In this scenario, QL performance is inferior to that of other controllers, yet all controllers are successful. The final simulated flight includes model parameter uncertainties, and the FQL attitude tracking error is better than the other two. Although altitude tracking error and control effort of DI are superior, the differences are insignificant.
\begin{table}[h!]
	\caption{Attitude tracking error, altitude tracking error, and control effort of three methods in different flight conditions. \label{tab6}}
\resizebox{\textwidth}{!}{%
	\begin{tabular}{ccccc}
		\hline
		\textbf{Flight Condition} & \textbf{Control Method}  & \begin{tabular}{c}
			\textbf{Attitude Tracking}\\ \textbf{Error (deg)}	
		\end{tabular}& \begin{tabular}{c}\textbf{Altitude Tracking} \\ \textbf{Error (m)}\end{tabular}& \begin{tabular}{c}
		\textbf{Control Effort}\\ \textbf{(deg)}
	\end{tabular}\\
		\hline
		 & Dynamic Inv & 0.057 & 0.662 & \textbf{0.593} \\
		Ideal & Q-learning & 0.047 & \textbf{0.643} & 0.649 \\
		 & Fuzzy QL & \textbf{0.040} & 0.644 & 0.669 \\
		 \hline
		 & Dynamic Inv  & \textbf{2.541} & \textbf{1.477} & 13.772\\
		Noise + Disturbance	& Q-learning & 11.178 & 21.655 &  \textbf{1.998} \\
	    & Fuzzy QL & 2.629 & 2.17 & 3.954  \\
	    \hline
	    & Dynamic Inv & 0.075 & 0.717 & \textbf{0.592} \\
	    Actuator Fault & Q-learning & 0.091 & 0.755 & 0.624\\
	    & Fuzzy QL & \textbf{0.064} & \textbf{0.708} & 0.654\\
	    \hline
	    & Dynamic Inv & 0.060 & \textbf{0.903} & \textbf{0.734}\\
	    Model Uncertainty & Q-learning & 0.056 & 0.936 & 0.793\\
	    & Fuzzy QL & \textbf{0.042} & 0.928 & 0.812 \\
		\hline
	\end{tabular}}
\end{table}
Another discussion is around the working area of the proposed FQL controller to prove its robustness. Although the previous simulations were in different flight conditions, they were designed just around one working point. More precisely, they were simulated in initial aircraft longitudinal speed $160\frac{m}{s}$, and constant aircraft longitudinal aerodynamic and control coefficients. In this part, the performance of both FQL and DI controllers are examined in a wider alteration of aforesaid coefficients and aircraft speed. In this case, the coefficients are varied between $\pm30\%$, and also the aircraft speed initializes between $150$ to $220\frac{m}{s}$. The FQL findings were satisfactory in comparison with one of the well-known robust controllers namely Dynamic Inversion. According to figure \ref{fig:11}, the FQL pitch angle tracking error varies between $0.04$ to $0.07$ degrees during $81$ defined initial conditions. Correspondingly, the control effort related to this controller varies between 0.64 to 2.6 degrees. On the other hand, the tracking error of DI computed between 0.05 to 0.07 degrees where the elevator control effort is varied from 0.7 to 2,6 degrees.
\begin{figure}[h!]
	\begin{center}
		\includegraphics[width=14.5cm]{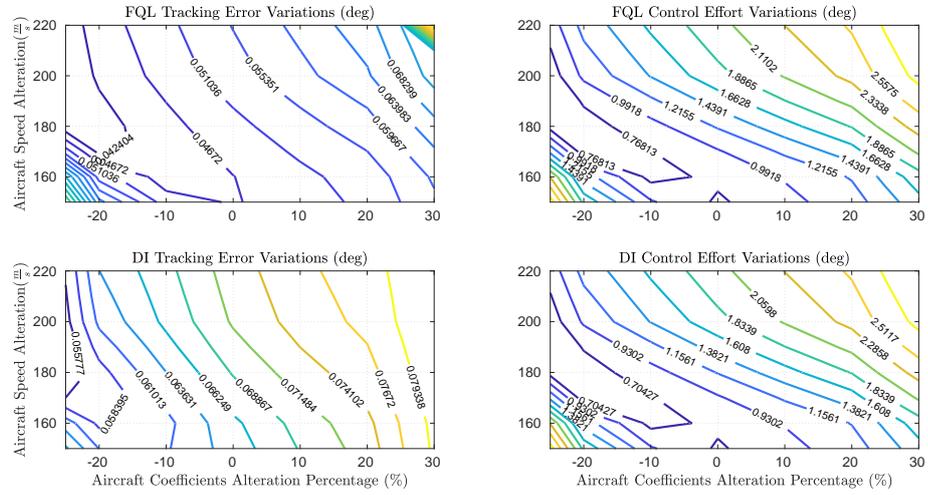}
		\caption{Fuzzy Q-learning and Dynamic Inversion tracking error and control effort examined by alteration of model parameters, and aircraft speeds}
		\label{fig:12}
	\end{center}
\end{figure}
\section{Conclusion}
In this research, auto-landing control of a regional jet aircraft with a novel configuration and specific longitudinal dynamic stability characteristics was addressed using Fuzzy Q-learning. The robustness property of this method was evaluated in several probable scenarios including atmospheric disturbances, sensor measurement noise, actuator faults, and model parameter uncertainties. The simulation results illustrated comparable improvements in contrast to Dynamic Inversion and classic Q-learning controllers. An innovative continuous action generator was proposed in this research to be a connector between optimal Q-tables and RL environments. In order to depict the robustness and working area of the proposed method, the aircraft's longitudinal speed, and coefficients have been altered widely, and the pitch angle tracking error as well as control effort are reported numerically. Summing up, the competency of the Fuzzy Q-learning method was proved in this problem approaching different flight conditions without being stuck in complicated Artificial Neural Network architectures.

\newpage
\bibliographystyle{unsrt}
\bibliography{ref}
\end{document}